# Quantification of the atomic surfaces and volumes of a metal cluster based on the molecular surface model

Yifan Yu[a,b,c,*], Junzhi Cui [b,c]

[a]Information Research Center of Military Science, PLA Academy of Military Science, Beijing 100142, China

[b]LSEC, ICMSEC, Academy of Mathematics and Systems Science, Chinese Academy of Sciences, Beijing 100190, China

[c]School of Mathematical Sciences, University of Chinese Academy of Sciences, Beijing 100049, China

**\*Corresponding author**

E-mail address: yuyifan@lsec.cc.ac.cn (Yifan Yu).

**Abstract**

The atomic volume and surface are important geometric quantities for calculating various macroscopic physical quantities from atomic models. This paper proposes a new analytical method to calculate the atomic volumes and surfaces of a metal cluster. This method adopts metallic radii to describe atom sizes and constructs the overall volume/surface by the molecular surface (MS) model. It divides cluster atoms into two types: interior atoms and boundary atoms. For an interior atom, the method defines a variational Voronoi cell as its volume. For a boundary atom, the method defines the intersection of the overall cluster volume and its variational Voronoi cell as its volume. The atomic surfaces are calculated along with the volume calculations. This new method considers the effect of atom sizes and can calculate not only the overall volume of a cluster but also the individual volume for each atom. This method provides computational support for multiscale coupled calculations from the microscale to macroscale.

# 1 Introduction

Molecular dynamics (MD) is a powerful and widely used research tool that can help researchers study the dynamics of materials at the atomic scale. The result of a molecular dynamics simulation is the trajectory of a group of atoms. From the trajectory, many physical quantities can be calculated according to multiscale theory, such as stress and strain [1-5]. The atomic volume and surface area are two geometric quantities, and they play an important role in calculating the physical quantities of an atomic cluster. For example, to calculate the first Piola-Kirchhoff stress distribution within an atomic cluster, the intermediate quantities need to be divided by atomic volumes [2-4]. However, an individual atomic volume is not well defined or easy to compute in a MD simulation [6-10].

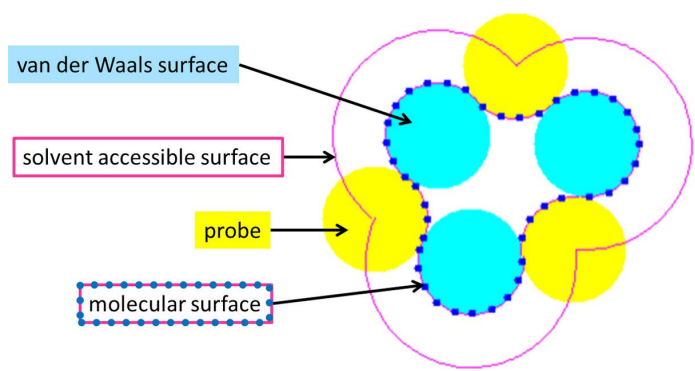

Fig. 1 The solvent accessible surface (SA), molecular surface (MS) and van der Waals surface (VW) models. [11]

Lee and Richards presented the three-dimensional shape representation models of solvent accessible surface (SA) and molecular surface (MS) for the protein [12]. The molecular volume obtained from the MS model is named the solvent excluded volume. In the special case of a point-sized solvent, these two types of molecular surfaces coincide, referred to as the van der Waals surface (VW). These surface models are shown in Fig. 1. The surface area and volume computational methods based on SA, MS, and VW models can be broadly divided into two types: approximation methods [13-22] and analytic methods [23-29]. Most approximation methods involve certain discretizations, such as polyhedral (generally triangular) decompositions or the representation of a surface with a large number of dots. PyMOL [30] and VMD [31] adopted approximation methods to generate the surface representation of a protein. Representing atoms by spherical balls provides an opportunity for analytically calculating the surface areas and volumes of an atom cluster. Connolly developed an analytical molecular surface and volume calculation method based on the MS model [23, 25]. McConkey introduced an analytic algorithm to quantify the atomic surface and volume for a protein based on the SA model [32]. McConkey defined an atomic volume as the intersection between the atom sphere and its Voronoi cell. Based on McConkey's work, Cazals et al. gave a derivation proof of the calculation formulae of the atomic surface and volume [33]. In addition, they implemented a computational program and performed a detailed numerical analysis.

The SA, MS, and VW models were originally designed for protein molecules without regard to metallic materials. One general way to compute individual atom volumes for metallic materials is calculating the Voronoi tessellation of atoms and defining the Voronoi cells as individual volumes [3, 4, 6, 7]. However, this method treats atoms as points without considering the effect of atom sizes, and it is sometimes wrong for boundary atoms because the Voronoi cells of the boundary atoms may be unbounded. Stukowski [34] proposed a geometric model to characterize the surface and volume of a metal cluster. First, Stukowski performed Delaunay triangulation to a metal cluster and obtained a set of tetrahedrons. Then, to identify the details of the solid surface and the holes or small gaps in the cluster, Stukowski introduced the concept of a probe sphere. Stukowski chose the size of the probe sphere, compared the probe sphere with the circumscribed sphere of each tetrahedron in the space, and then deleted the tetrahedrons whose circumscribed sphere radius was larger than the probe sphere. The remaining tetrahedrons were the metal cluster volume, and the outer faces of these tetrahedrons formed the cluster surface. However, this model also treats atoms as points without considering the atom size effect, and it can only calculate the overall volume/surface of a metal cluster, not the individual volume/surface for each atom.

In this paper, we propose a new method for calculating the atomic surfaces and volumes of a metal cluster. This new method is mainly based on the MS model and absorbs some of the features of Stukoiski's model. First, this method adopts metallic radii to describe atom sizes and constructs the overall volume/surface by the MS model. Then, it divides cluster atoms into two types: interior atoms and boundary atoms. For an interior atom, the method defines a variational Voronoi cell as its volume. For a boundary atom, the method defines the intersection of the overall cluster volume and its variational Voronoi cell as its volume. The atomic surfaces are calculated along with calculating the volumes. Essentially, we define all atomic volumes (both interior and boundary atoms) as the intersection of the overall cluster volume and the corresponding Voronoi cells. Here, we do not adopt the atomic volume definition by McConkey [32] and Cazals [33]: the atom sphere intersects with its Voronoi region, because their method was based on the SA model, while ours is based on the MS model. Our new atomic volume/surface method takes into account the effect of atom sizes and can calculate not only the overall volume/surface of a cluster but also the individual volume/surface for each atom.

The structure of this paper is as follows. Section 2 reviews the MS model and surface definition equations. Section 3 gives the surface calculation methods for a metal cluster and individual atoms. Section 4 gives the corresponding volume calculation methods. Section 5 discusses the self-intersecting surface problem encountered in constructing a MS surface. Section 6 discusses the probe radius selection problem. Section 7 gives numerical examples of 3D copper (Cu) nanowire stretching and bending to verify the effectiveness of the volume/surface calculation method. In Section 8, the conclusion is presented.

**2 Review of the MS model and surface definition equations**

The molecular surface (MS) model defines a geometric description of the outer and inner boundaries of an atom cluster. The molecular surface is a smooth network of convex and reentrant faces traced by the inward-facing part of a probe sphere as it rolls over the cluster [23]. The volume obtained by the MS model is also called the solvent-excluded volume [25]. It is a spatial region from which the solvent is excluded by the presence of the molecule (atom cluster). The solvent is also modeled as a hard sphere and chosen as the probe sphere. The solvent-excluded volume is the van der Waals volume plus the interstitial volume. The interstitial consists of packing defects between atoms that are too small to admit a probe sphere. This means that the MS model can characterize defect structures in a metal cluster if an appropriate probe sphere is selected. Therefore, the MS model is suitable for describing the occupied volume of a metal cluster. Another advantage of the MS model is that the surface it generates is smooth, which facilitates subsequent multiscale calculations of physical and chemical properties for an atom cluster.

The MS definition equations are given in Table 1. A torus between atoms $i$ and $j$ is generated as a probe sphere rolls around the pair of atoms. The torus has an axis $\mathbf{u}_{ij}$, center $\mathbf{t}_{ij}$, and radius $r_{ij}$. The circle of contact between atom $i$ and the probe sphere has a center $\mathbf{c}_{ij}$, radius $r_c$, and signed displacement $d_c$ from the center of atom $i$ to center $\mathbf{c}_{ij}$. A tetrahedron is generated when the probe is simultaneously tangent to three atoms $i$, $j$, and $k$. Its vertices are the centers of the probe and atoms $i$, $j$, and $k$, and the base triangle connects the three atom centers. $\mathbf{P}_{ijk}$ is the center of the probe and is at a height $h_{ijk}$ above a base point $\mathbf{b}_{ijk}$ lying on the base triangle. $v_{pi}$ is a vertex of a concave triangle, which is the contact point between the probe and atom $i$. Concave triangles meet convex faces at those vertices. $\beta_v$ and $\alpha_v$ are the interior angles of a concave triangle and convex face, respectively. $\varphi_s$ is the angle of a saddle face wrapping around the torus axis. $\theta_{si}$ is the saddle-width angle of a saddle face bordering a convex edge on atom $i$.

Table 1 Surface definition equations [23, 25].

| Variable name | Value |
|---|---|
| Atomic coordinate | $\mathbf{a}_i, \mathbf{a}_j, \mathbf{a}_k, \ldots$ (input) |
| van der Waals radii | $r_i, r_j, r_k, \ldots$ (input) |
| Probe radius | $r_p$ (input) |
| Interatomic distance | $d_{ij} = |\mathbf{a}_j - \mathbf{a}_i|$ |
| Torus axis unit vector | $\mathbf{u}_{ij} = (\mathbf{a}_j - \mathbf{a}_i)/d_{ij}$ |
| Torus center | $\mathbf{t}_{ij} = \frac{1}{2}(\mathbf{a}_i + \mathbf{a}_j) + \frac{1}{2}(\mathbf{a}_j - \mathbf{a}_i)$ $\times \left[ (r_i + r_p)^2 - (r_j + r_p)^2 \right]/d_{ij}^2$ |
| Torus radius | $r_{ij} = \frac{1}{2}\left[ (r_i + r_j + 2r_p)^2 - d_{ij}^2 \right]^{1/2}$ $\times \left[ d_{ij}^2 - (r_i - r_j)^2 \right]^{1/2}/d_{ij}$ |
| Base triangle angle | $\omega_{ijk} = \arccos(\mathbf{u}_{ij}, \mathbf{u}_{ik})$ |
| Base plane normal vector | $\mathbf{u}_{ijk} = \mathbf{u}_{ij} \times \mathbf{u}_{ik}/\sin\omega_{ijk}$ |
| Torus-basepoint unit vector | $\mathbf{u}_{tb} = \mathbf{u}_{ijk} \times \mathbf{u}_{ij}$ |
| Base point | $\mathbf{b}_{ijk} = \mathbf{t}_{ij} + \mathbf{u}_{tb}\left[ \mathbf{u}_{ik} \cdot (\mathbf{t}_{ik} - \mathbf{t}_{ij}) \right]$ $\times (\sin\omega_{ijk})^{-1}$ |
| Probe height | $h_{ijk} = \left[ (r_i + r_p)^2 - |\mathbf{b}_{ijk} - \mathbf{a}_i|^2 \right]^{1/2}$ |
| Probe position | $\mathbf{p}_{ijk} = \mathbf{b}_{ijk} \pm h_{ijk}\mathbf{u}_{ijk}$ |
| Vertex | $\mathbf{v}_{pi} = (r_i\mathbf{p}_{ijk} + r_p\mathbf{a}_i)/(r_i + r_p)$ |
| Contact circle center | $\mathbf{c}_{ij} = (r_i\mathbf{t}_{ij} + r_p\mathbf{a}_i)/(r_i + r_p)$ |
| Contact circle radius | $r_c = r_{ij}r_i/(r_i + r_p)$ |
| Contact circle displacement | $d_c = \mathbf{u}_{ij} \cdot (\mathbf{c}_{ij} + \mathbf{a}_i)$ |
| Concave arc plane normal vector | $\mathbf{n}_{ijk} = (\mathbf{p}_{ijk} - \mathbf{t}_{ij}) \times \mathbf{u}_{ij}/r_{ij}$ |
| Concave triangle angle | $\beta_v = \arccos(\mathbf{n}_{ijk} \cdot \mathbf{n}_{ikj})$ |

| Convex face angle | $\alpha_v = \pi - \beta_v$ |
|---|---|
| Saddle wrap angle | $\varphi_s = \arccos(\mathbf{n}_{ijk} \cdot \mathbf{n}_{ijl})$ when $\mathbf{n}_{ijk} \times \mathbf{n}_{ijl} \cdot \mathbf{u}_{ij} \geq 0$ $= -\arccos(\mathbf{n}_{ijk} \cdot \mathbf{n}_{ijl}) + 2\pi$ when $\mathbf{n}_{ijk} \times \mathbf{n}_{ijl} \cdot \mathbf{u}_{ij} < 0$ |
| Saddle width angle | $\theta_{si} = \arctan(d_c / r_c)$ |
| Euler characteristic | $\chi = 2 - $ number of cycles |

## 3 Surface area calculations

Based on the MS model, we can generate an outer surface for an atom cluster, which makes the concept of atom cluster volume meaningful.

### 3.1 Metal cluster surface area

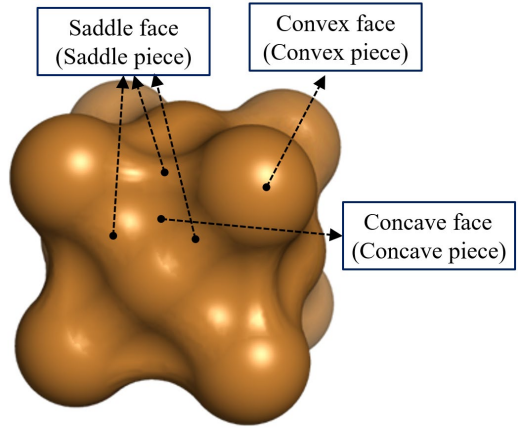

Fig. 2 The surface of an atom cluster (FCC unit cell).

The surface of an atom cluster consists of three types of faces: concave, saddle and convex faces, as shown in Fig. 2. The surface area of an atom cluster is

$$A_C = \sum_{c^-} A_{c^-} + \sum_s A_s + \sum_{c^+} A_{c^+}, \quad (1)$$

where $A_{c^-}$ is a concave face area, $A_s$ is a saddle face area, and $A_{c^+}$ is a convex face area. The face areas of those three types can be calculated by using the radii and angles presented in Table 1.

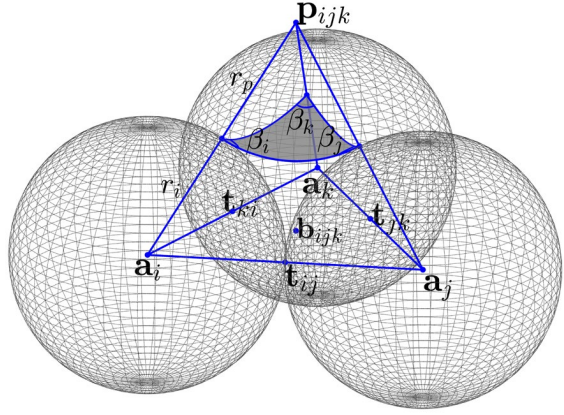

Fig. 3 A concave face (shaded).

Figure 3 shows a concave face, which is a spherical triangle, and its boundary is three concave arcs. The area of a concave face is

$$A_{c^-} = r_p^2 \left( \sum_v \beta_v - \pi \right), v \in \{i, j, k\}, \qquad (2)$$

where $r_p$ is the probe radius and $\beta_v$ is a concave triangle angle.

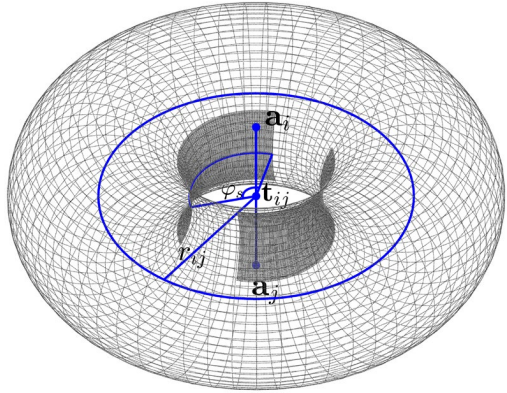

Fig. 4 Two saddle-shaped faces (shaded).

Figure 4 shows two saddle faces. A saddle face is part of a torus surface, and its boundary is convex and concave arcs. The area of a saddle face is

$$A_s = \varphi_s \left[ r_{ij} r_p \left( \theta_{si} + \theta_{sj} \right) - r_p^2 \left( \sin \theta_{si} + \sin \theta_{sj} \right) \right], \qquad (3)$$

where $\varphi_s$ is a saddle wrap angle, $r_{ij}$ is a torus radius, and $\theta_{si}$ and $\theta_{sj}$ are saddle width angles.

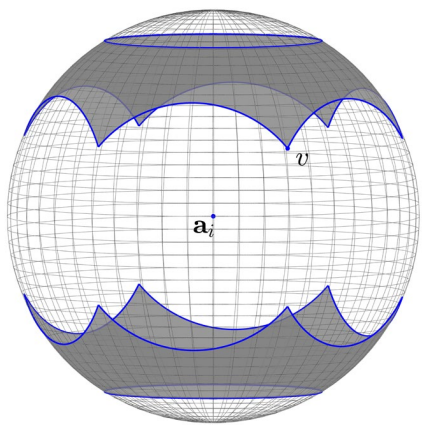

Fig. 5 Two convex faces (shaded).

Figure 5 shows two convex faces, and its boundary is cycles of convex arcs. Convex faces are generated for each partially or wholly accessible atom. Buried atoms generate no surface. The area of a convex face is

$$A_{c^+} = r_i^2 \left[ 2\pi\chi - \sum_s \phi_s \sin\theta_{si} - \sum_v (\pi - \alpha_v) \right], \qquad (4)$$

where $r_i$ is an atom radius and $\alpha_v$ is a convex angle. Figure 6 shows the relationship between $\alpha_v$ and $\beta_v$ at vertex $v$.

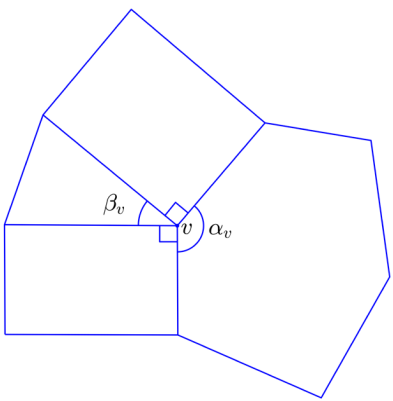

Fig. 6 The two saddle rectangles (right angles), one concave triangle ($\beta_v$) and one convex region ($\alpha_v$) meet at a vertex $v$. The sum of the angles is $2\pi$.

## 3.2 Atomic surface areas of a metal cluster

Here, an atomic surface refers to the surface of an atom occupied volume. There are two types of atoms in a cluster: boundary atoms and interior atoms. Here, a variational Voronoi cell is defined. Based on it and the MS model, we can define the individual atomic surfaces for both boundary atoms and interior atoms.

## The surface area of an interior atom

As mentioned above, the normal Voronoi cell is not suitable as the occupied volume of an interior atom. We define a variational Voronoi cell of atom $i$ using a new distance instead of the Euclidean distance

$$R_i = \{\mathbf{x} \mid d(\mathbf{x},\mathbf{a}_i) \leq d(\mathbf{x},\mathbf{a}_j) \text{ for all } j \text{ that generates a torus with } i\}, \tag{5}$$

$$d(\mathbf{x},\mathbf{a}_i) \leq d(\mathbf{x},\mathbf{a}_j) \Leftrightarrow \frac{(\mathbf{x}-\mathbf{a}_i)\cdot(\mathbf{t}_{ij}-\mathbf{a}_i)}{|\mathbf{t}_{ij}-\mathbf{a}_i|^2} \leq \frac{(\mathbf{x}-\mathbf{a}_j)\cdot(\mathbf{t}_{ij}-\mathbf{a}_j)}{|\mathbf{t}_{ij}-\mathbf{a}_j|^2}, \tag{6}$$

where the faces of that variational Voronoi cell are planes perpendicular to the lines between atom $i$ and its adjacent atoms, and the planes pass through the points $\mathbf{t}_{ij}$ on the lines.

Figure 7 shows Voronoi tessellations of a cluster, all interior Voronoi cells are colored. Atom $i$ is located in the center, its radius is smaller than other atoms, and the radii of other atoms are the same. Figure 7(a) shows the normal Voronoi tessellation using the Euclidean distance, all interior Voronoi cells are the same size. Figure 7(b) shows the variational Voronoi tessellation using distance Eqs. (5)(6), and the Voronoi cell of atom $i$ is smaller than other interior Voronoi cells. This means that variational Voronoi cells can reflect the influence of atom sizes. From the perspective of atomic physics, the larger an atom is, the larger the region it affects. Therefore, the variational Voronoi cell is more suitable as the interior atom occupied volume than the normal cell.

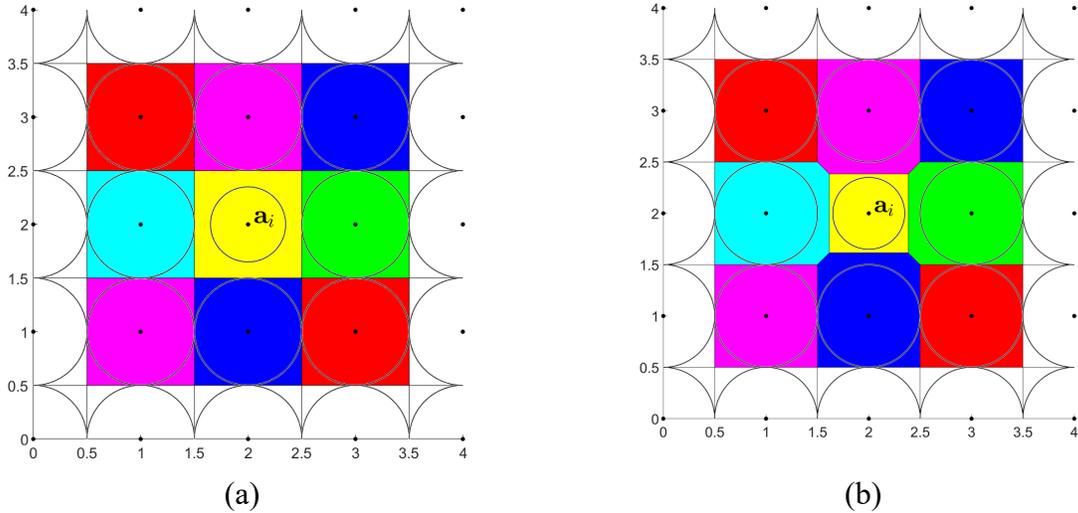

Fig. 7 Voronoi tessellations of an atom cluster (circles represent atom spheres and points represent atom centers). (a) Normal Voronoi cells (colored regions represent cells of interior atoms) and (b) variational Voronoi cells.

The faces of a variational Voronoi cell are polygons. The surface area of the cell is the sum of those polygonal areas that can be calculated using the vertices of the cell. There are two ways to describe a variational Voronoi cell: one is by a set of inequalities, and the other is by vertex coordinates. Eqs. (5) and (6) describe a variational Voronoi cell by a set of inequalities. Through the primal–dual methods [35, 36], the inequalities can be transformed into a set of vertices of the Voronoi cell.

*The surface area of a boundary atom*

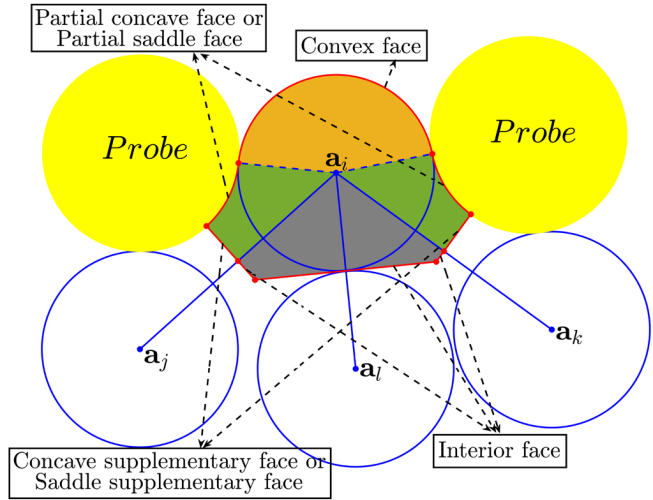

Fig. 8 The surface of a boundary atom (dark yellow, convex volume; dark green, partial concave volume or partial saddle volume; gray, atomic interior volume).

As shown in Fig. 8, the surface of a boundary atom consists of six parts: partial concave faces, partial saddle faces, convex faces, interior faces, concave supplementary faces and saddle supplementary faces. The surface area of atom $i$ is

$$A_i = \sum_{c_i^-} A_{c_i^-} + \sum_{s_i} A_{s_i} + \sum_{c_i^+} A_{c_i^+} + \sum_{n_i} A_{n_i} + \sum_{\tilde{c}_i} A_{\tilde{c}_i} + \sum_{\tilde{s}_i} A_{\tilde{s}_i},$$

where $c_i^-$, $s_i$, $c_i^+$, $n_i$, $\tilde{c}_i$, and $\tilde{s}_i$ represent a partial concave face, partial saddle face, interior face, concave supplementary face, and saddle supplementary face, respectively.

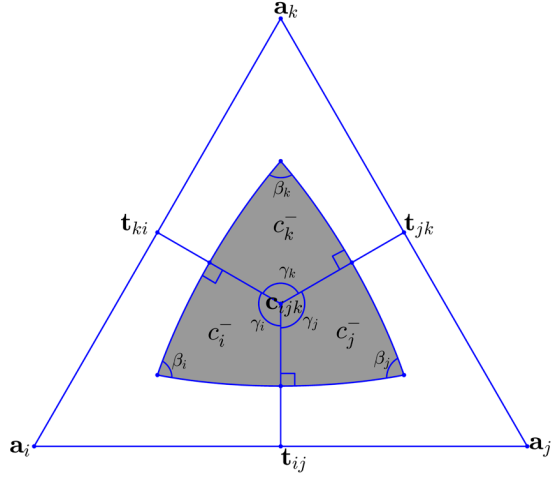

Fig. 9 The vertical view of a concave face.

A concave face is shared by three atoms, as shown in Fig. 9. It should be divided into three parts. Each partial concave face belongs to the corresponding atom. There is one torus between every two atoms. Three tori converge at the concave face. The middle section planes of those tours intersect at point $\mathbf{c}_{ijk}$ on the concave face, and they divide the concave face into three parts, $c_i^-$, $c_j^-$ and $c_k^-$, each of which corresponds to atoms $i$, $j$, and $k$, respectively. The area of partial concave face $c_i^-$ is

$$A_{c_i^-} = r_p^2 \cdot (\beta_i + \gamma_i + \pi), \tag{7}$$

where $\beta_i$ is a concave triangle angle corresponding to atom $i$, and $\gamma_i$ is an interior angle on $\mathbf{c}_{ijk}$.

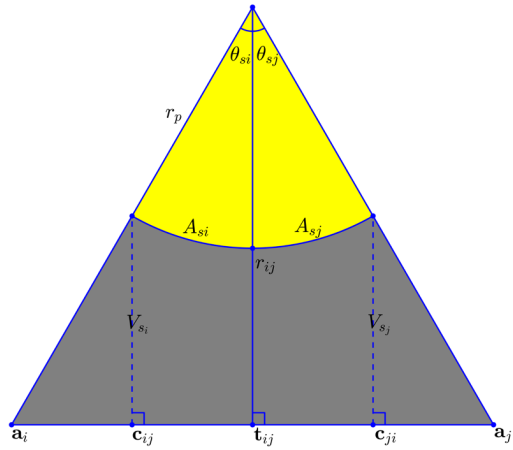

Fig. 10 A saddle face between atoms $i$ and $j$.

A saddle face is shared by two atoms, as shown in Fig. 10. It should be divided into two parts. Each partial saddle face belongs to the corresponding atom. As shown in Fig. 10, a saddle face is a part of the torus that is between atoms $i$ and $j$. The middle section of the torus divides the saddle face into two parts that belong to atoms $i$ and $j$, denoted as partial saddle faces $s_i$ and $s_j$, respectively. The area of $s_i$ is

$$A_{s_i} = \varphi_s r_p \left( r_{ij} \theta_{si} - r_p \sin \theta_{si} \right), \tag{8}$$

where $\varphi_s$ is a saddle wrap angle, $r_{ij}$ is a torus radius, and $\theta_{si}$ and $\theta_{sj}$ are saddle width angles.

As shown in Fig. 8, a convex face only relates to a single atom, so its area is completely contained in that atom surface area. A convex face area is given by Eq. (4).

As shown in Fig. 8, interior faces are the red lines of the gray region, which are inside the cluster. In fact, an interior face is a part of the variational Voronoi partition plane between two atoms. Assuming that atom $i$ forms a torus with atom $j$, the inequalities describing the shaded region of atom $i$ are

$$\begin{cases} \vec{\mathbf{x}} \cdot \vec{\mathbf{v}_{ij}} \leq \vec{\mathbf{t}_{ij}} \cdot \vec{\mathbf{v}_{ij}} & \text{for all } j, \\ \vec{\mathbf{x}} \cdot \vec{\mathbf{n}_h} \leq \vec{\mathbf{a}_i} \cdot \vec{\mathbf{n}_h} & \text{for all } T_h, \end{cases} \tag{9}$$

where $\mathbf{x}$ is a point in the shaded region, $\mathbf{t}_{ij}$ is a torus center between $\mathbf{a}_i$ and $\mathbf{a}_j$, $v_{ij}$ is the unit direction vector from $\mathbf{a}_i$ to $\mathbf{a}_j$, $T_h$ is a triangle formed by three boundary atoms (including atom $i$), and $\mathbf{n}_h$ is an outward unit normal vector of $T_h$. Similar to calculating the face areas of an interior atom variational Voronoi cell above, the primal–dual methods [35, 36] can be used to convert equalities of Eq. (9) to vertices of the shaded region. Then, interior face areas can be obtained.

For a boundary atom, each related concave face produces two supplementary faces, called concave supplementary faces. Figure 11 shows concave supplementary faces as shaded areas. A concave supplementary face is the intersection of a torus middle section and the region between the concave face and triangle base. In fact, it lies on an unbound face of the Voronoi cell of a boundary atom. The concave supplementary face area $A_{\bar{c}_i}$ is

$$A_{\bar{c}_i} = A_{\bar{c}_{ij}} + A_{\bar{c}_{ik}}, \tag{10}$$

$$A_{\bar{c}_{ij}} = \left( h_{ijk}/2 \right) \left[ (r_i + r_p)^2 - \| \mathbf{t}_{ij} - \mathbf{a}_i \|^2 - h_{ijk}^2 \right]^{1/2} - (r_p^2/2) \times \arccos \left( h_{ijk} / \left[ (r_i + r_p)^2 - \| \mathbf{t}_{ij} - \mathbf{a}_i \|^2 \right]^{1/2} \right), \tag{11}$$

where $A_{\bar{c}_{ik}}$ and $A_{\bar{c}_{ij}}$ are calculated in the same way (a triangle minus a sector), and the formula for calculating $A_{\bar{c}_{ik}}$ can be obtained by simply transforming Eq. (11).

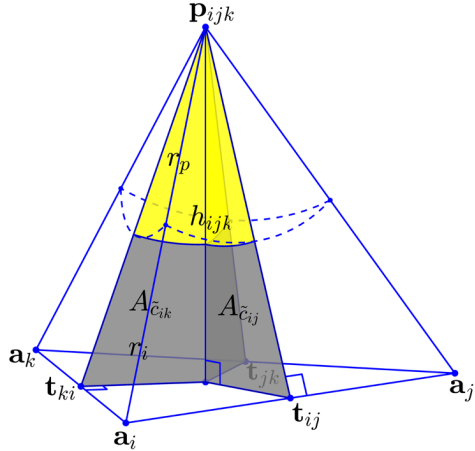

Fig. 11 Concave supplementary faces.

Each saddle face produces a supplementary surface called the saddle supplementary face. A saddle supplementary face is the intersection of a saddle piece (as defined hereinafter), a torus middle section and the corresponding unbounded face of the variational Voronoi cell, as shown in Fig. 12. The saddle supplementary face area $A_{\tilde{s}_i}$ is

$$A_{\tilde{s}_i} = \frac{1}{2}\varphi_s (r_{ij} - r_p)^2, \qquad (12)$$

where $\varphi_s$ is a saddle wrap angle and $r_{ij}$ is a torus radius.

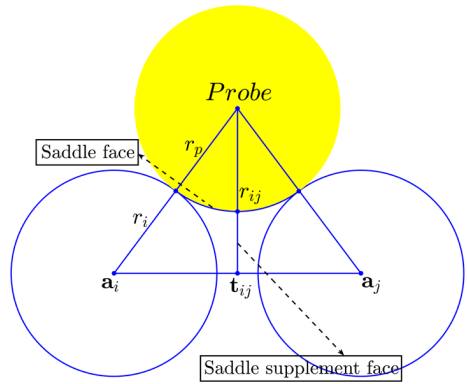

Fig. 12 A saddle supplementary face.

## 4 Volume calculations

According to the MS model definition and face area equations, we can analytically calculate the MS volume.

### 4.1 Metal cluster volume

As shown in Fig. 2, the MS volume of a cluster can be divided into four parts: interior polyhedron, convex pieces, saddle pieces, and concave pieces. A cluster volume is

$$V_C = V_n + \sum_{c^+} V_{c^+} + \sum_{s} V_s + \sum_{c^-} V_{c^-}, \tag{13}$$

where $V_n$ is the interior polyhedron volume, $V_{c^+}$ is a convex piece volume, $V_s$ is a saddle piece volume, and $V_{c^-}$ is a concave piece volume.

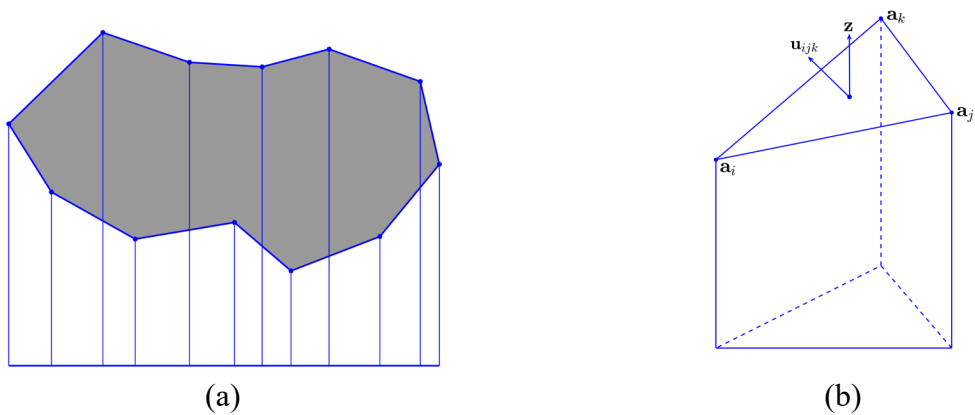

(a)            (b)

Fig. 13 The interior polyhedron. (a) Two-dimensional representation of an interior polyhedron (shaded) and (b) truncated triangular prism.

The interior polyhedron is inside the molecular surface, and it can be derived from the molecular surface. For each concave triangle of the surface, a flat triangle is constructed between the centers of the three atoms that the concave triangle bridges. Those flat triangles define the interior polyhedron. The volume of the interior polyhedron is

$$A_f = \tfrac{1}{2} d_{ij} d_{ik} \sin \omega_{ijk}, \tag{14}$$

$$V_n = \sum_f \tfrac{1}{3} \mathbf{z} \cdot (\mathbf{a}_i + \mathbf{a}_j + \mathbf{a}_k)(\mathbf{u}_{ijk} \cdot \mathbf{z}) A_f, \tag{15}$$

where $A_f$ is the area of a base triangle between the centers of atoms $i$, $j$ and $k$, $\mathbf{u}_{ijk}$ is a triangle face normal vector, and $\mathbf{z}$ is a unit vector pointing along the positive z-axis.

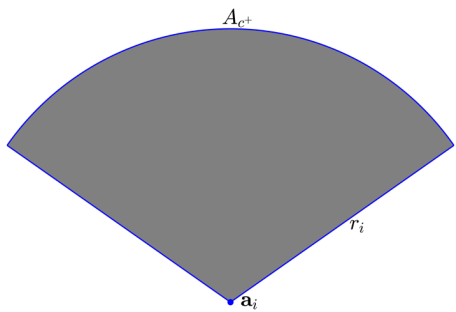

Fig. 14 A convex piece.

A convex piece is the volume between the center of an atom and its convex face (Fig. 14). The volume of a convex piece is

$$V_{c^+} = \tfrac{1}{3} r_i A_{c^+}, \tag{16}$$

where $A_{c^+}$ is the convex face area.

A saddle piece (shaded part in Fig. 10) can be divided into four parts. Between the center of each atom and its circle of contact with the probe sphere is a sector of cone. Between these two conical pieces is the volume of a sector of hole inside the torus, which for ease of later computing the atom occupied volume is divided into two parts lying on either side of the plane bisecting the torus. The volume of a saddle piece is

$$V_s = V_{ci} + V_{si} + V_{sj} + V_{cj}, \tag{17}$$

$$V_{ci} = (\phi_s / 6) r_i^3 \sin\theta_{si} \cos^2\theta_{si}, \tag{18}$$

$$V_{si} = (\phi_s / 2)\left[ r_{ij}^2 r_p \sin\theta_{si} - r_{ij} r_p^2 (\sin\theta_{si}\cos\theta_{si} + \theta_{si}) + (r_p^3/3) \times \\ (\sin\theta_{si}\cos^2\theta_{si} + 2\sin\theta_{si}) \right]. \tag{19}$$

where $V_{sj}$ and $V_{cj}$ can be obtained by simply transforming Eq. (18)(19).

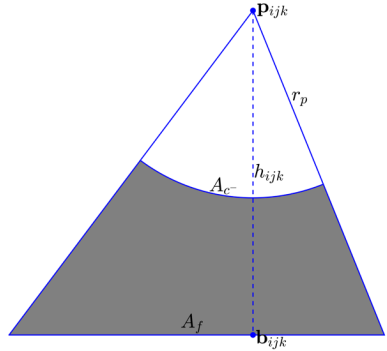

Fig. 15 A concave piece.

A concave piece is the volume of a triangular pyramid minus the volume of a piece of the probe sphere, shown in Fig. 15. The volume of a concave piece is

$$V_{c^-} = \tfrac{1}{3}\left( h_{ljk} A_f - r_p A_{c^-} \right), \tag{20}$$

where $A_{c^-}$ is the concave face area.

### 4.2 Atomic volumes of a metal cluster

As mentioned above, there are two types of atoms in the cluster: boundary atoms and interior atoms. Different types of atoms calculate their volumes in different ways.

*The volume of an interior atom*

As with the surface area calculations above, the variational Voronoi cell is regarded as the interior atom occupied volume. The vertices of the variational Voronoi cell can be obtained by Eqs. (5)(6) and the primal–dual methods [35, 36]. The volume of a variational Voronoi cell can be computed by dividing it into congruent pyramids. Each pyramid has a face of the polyhedron as its base and the center of the polyhedron as its apex.

Therefore, the Voronoi cell volume for an interior atom $i$ is

$$V_i = \sum_j A_{F_j} d_j / 3, \quad (21)$$

where $A_{F_j}$ is a face area of the Voronoi cell and $d_j$ is the distance from the center of the cell to face $F_j$.

### *The volume of a boundary atom*

The normal Voronoi cell of a boundary atom may be unbounded, and thus it cannot represent the occupied volume of a boundary atom. The surface of a boundary atom we defined above encloses a bounded region, and the larger the atom is, the larger the enclosed region. This enclosed region can be regarded as the occupied volume of a boundary atom.

As shown in Fig. 8, the volume of a boundary atom is divided into four parts

$$V_i = V_{n_i} + \sum_{c_i^+} V_{c_i^+} + \sum_{s_i} V_{s_i} + \sum_{c_i^-} V_{c_i^-}, \quad (22)$$

where $V_{n_i}$ is the atomic interior volume, $V_{c_i^+}$ is a convex volume, $V_{s_i}$ is a partial saddle volume, and $V_{c_i^-}$ is a partial concave volume.

The atomic interior volume of a boundary atom (Fig. 8) is the intersection of its variational Voronoi cell and the interior polyhedron of the cluster. We mark the triangles on the surface of the interior polyhedron, which have that boundary atom as a vertex, and then use those triangles as the separating planes to obtain the inequalities toward the interior polyhedron

$$\mathbf{a}_{v_t} \cdot \mathbf{x} \leq b_{v_t} \quad 1 \leq t \leq n_t, \quad (23)$$

where $n_t$ represents the number of marked triangles. We obtain the inequalities of the interior part of a boundary atom by combining Eq. (9) and Eq. (23). Through the primal–dual methods [35, 36], we calculate the vertices of the interior part and further calculate its volume, denoted as $V_{n_i}$.

The convex piece is only related to an atom (Fig. 8). Therefore, the corresponding volume can simply be assigned to that atom. The volume formula is given by Eq. (16).

A partial saddle volume is a part of the volume of a saddle piece. A saddle piece (Fig. 8 and Fig. 10) is relevant to two atoms. A saddle piece volume should be split into two parts according to a certain rule. Here, we split the saddle piece by the middle section plane of the torus between the two atoms. Each part volume is assigned to the corresponding atom. For a torus piece between atoms $i$ and $j$, a partial saddle volume for atom $i$ is

$$V_{s_i} = V_{ci} + V_{si}, \tag{24}$$

where $V_{ci}$ and $V_{si}$ are given by Eqs. (18)(19).

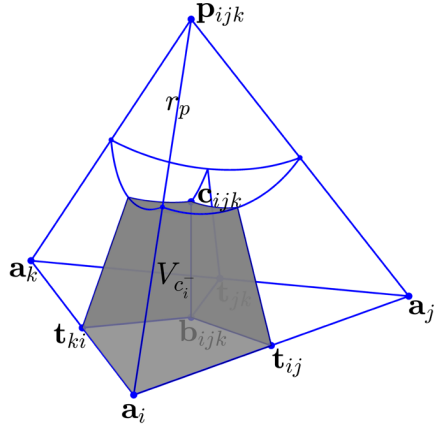

Fig. 16 A partial concave volume.

A partial concave volume is a part of the volume of a concave piece, as shown in Fig. 8 and Fig. 16. A concave piece is relevant to three atoms. The concave piece volume can be divided into three parts, and each part volume is assigned to the corresponding atoms. A concave face corresponds to a base triangle. The three middle sections of the tori related to the concave face divide the concave and base triangle into three parts, respectively. The intersections of three middle sections on the concave face and base triangle are points $c_{ijk}$ and $b_{ijk}$. Therefore, the concave piece is divided into three parts by the three middle sections, denoted as $V_{c_i^-}$, $V_{c_j^-}$, and $V_{c_k^-}$. For a concave piece generated by atoms $i$, $j$, and $k$, a partial concave volume for atom $i$ is

$$V_{c_i^-} = \tfrac{1}{3}\left(h_{ijk} A_{f_i} - r_p A_{c_i^-}\right), \tag{25}$$

where $A_{c_i^-}$ is given by Eq. (7).

## 5 Self-intersecting surface problem

A saddle face intersects itself when the torus radius is less than the probe radius, as shown in Fig. 17. It can also lead to two concave faces interpenetrating. Therefore, the calculation of the surface area and space occupied volume of the affected atoms must be wrong. These problems typically occur in deep grooves. The precise solution to this problem will be discussed in the next paper. At present, there is an approximate solution. The surface area and occupied volume of an affected atom is taken as the average of those of the neighboring unaffected atoms within a small range.

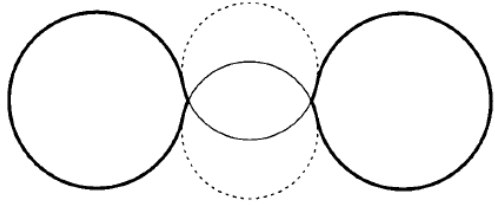

Fig. 17 Deep grooves in the MS model [37]. Two probe spheres coming from two opposite sides have a self-intersection.

## 6 Probe radius problem

The probe radius $r_p$ is artificially determined when constructing the surface area and volume of an atomic cluster. A larger radius makes the cluster surface smoother, and a smaller radius makes the surface present more details. Assuming that the radius of atom $i$ is greater than $j$, if $r_p$ is relatively large, the lateral mid-section plane of the tours between atoms $i$ and $j$ may directly pass through atom $j$. Obviously, we do not want that case to happen because the spherical volume of an atom itself should belong to the occupied region of that atom. Therefore, the section of $r_p$ needs to be carefully considered. In chemistry, the solvent molecule radius is often chosen as $r_p$ [32, 33, 37]. For a metal cluster, we generally choose the radius of the smallest atom in the cluster as $r_p$ because that selection can accurately characterize the metal cluster volume and ensure that the atom itself is completely located in the corresponding atom occupied region.

## 6 Numerical experiments

In continuum mechanics theory, stress is one of the most important physical quantities. How to correctly extract stress information from atomic simulations is a key to connecting micro/nanoscale mechanics and continuum mechanics. For a discrete atom system, the stress tensor for atom $i$ is given by

$$\sigma_{ab} = -\frac{mv_a v_b}{V} - \frac{W_{ab}}{V}, \qquad (26)$$

where $a$ and $b$ take on values $x$, $y$, $z$, and $V$ is the atomic occupied volume. The first term is the kinetic energy contribution for atom $i$, and the second term is the virial contribution due to intra- and intermolecular interactions, where the exact computation details are given in [6, 7].

In general, the per-atom volume in Eq. (26) is difficult to calculate. However, the method given by this paper could be applied to measuring the volume variation during energy minimizations and molecular dynamics simulations.

Here, we give two numerical experiments that use our volume calculation method to calculate stress distributions within configurations. It should be noted that to calculate accurate volumes, the prerequisite is to accurately calculate surface areas, as shown in the atomic volume calculation formulae above.

### *6.1 Cu nanowire stretching*

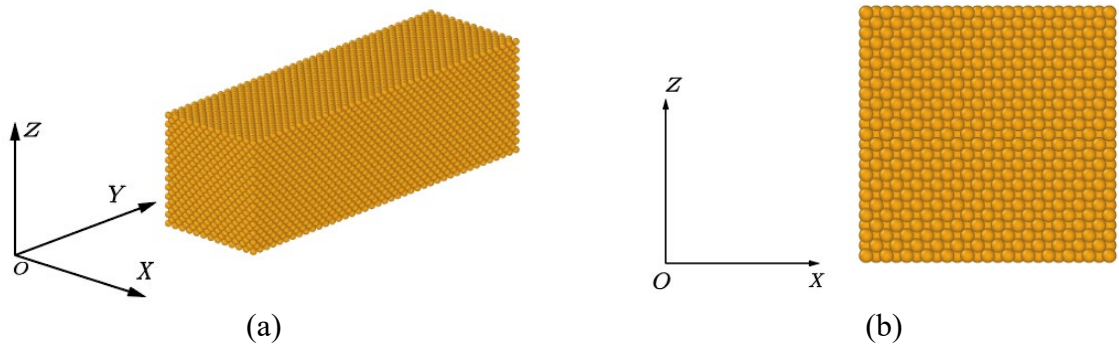

Fig. 18 Single-crystal Cu nanowire. (a) The initial configuration (43.38 Å × 144.6 Å × 43.38 Å) and (b) the cross-section (010)

Figure 18 shows the initial state of a Cu nanowire with the crystal direction [100] along the x-axis, [010] along the y-axis, and [001] along the z-axis. The Cu nanowire has 12×40×12 unit cells along the x, y and z directions, and its size is 43.38 Å×144.6 Å×43.38 Å for a total of 25,313 atoms. We loaded the Cu nanowire along the y-axis direction, and the x and z directions were set to have stress-free boundary conditions. We calculated the total potential energy of a Cu nanowire using the EAM potential function proposed in [38].

The loading was conducted by the quasi-static loading method. The nanowire was first relaxed for 60 ps to obtain its equilibrium configuration. For the relaxation method, the two layers of atoms at both ends were fixed, and all the remaining atoms were relaxed for 30 ps in the NVT (or canonical) ensemble. The target temperature was 300 K, and the temperature was controlled by the Nosé–Hoover thermostat [39]. Then, the last two layers of atoms in the negative y-axis direction were fixed, and the remaining atoms were moved certain distances along the y-axis direction in each loading step. The end atoms in the y-axis direction were moved by 0.2 Å, and the other atoms were moved in proportion to their relative locations. After each loading step, the nanowire was relaxed at a constant temperature of 300 K for 60 ps, and the average atomic positions in the last 2 ps were recorded.

Figure 19 shows the overall stress–strain curve of the single-crystal Cu nanowire during stretching. The stress of the 49th loading step passed the stress peak, indicating that defects were generated in the nanowire.

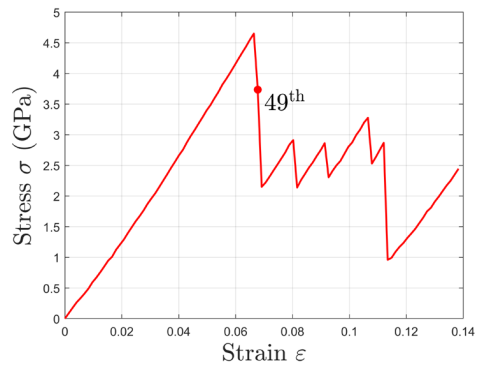

Fig. 19 The stretching stress–strain curve of the single-crystal Cu nanowire.

We chose the initial configuration and the 49th loading step configuration to calculate the stress distribution, and Figure 20 is the 49th loading step configuration.

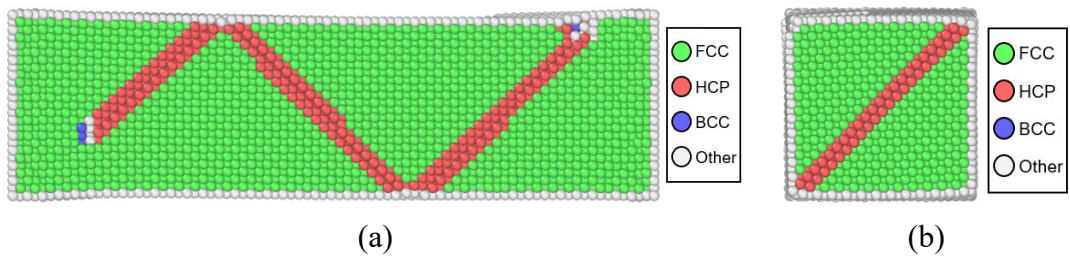

(a)  (b)

Fig. 20 The 49th loading step configuration. (a) The longitudinal section (x<0) and (b) the cross-section (y<0).

First, we calculated the atomic volumes of the initial configuration and 49th loading step configuration. Figure 21 shows the atomic volume distributions of the two configurations. In Fig. 21(a), atoms in the bin of [2.5, 3.5] were the edge atoms of the nanowire, atoms in the bin of [2.5, 3.5] were face atoms, and atoms in the bin of [10.5, 11.5] were interior atoms. In addition, each of 8 vertex atoms of the nanowire occupied a volume of 1.52 Å$^3$. Since their number was too small, they are not shown in the figure. In Fig. 21(b), the atomic volume distribution changed, and the occupied volumes of surface and interior atoms changed more obviously because the dislocations glided over the entire nanowire and the defects were generated.

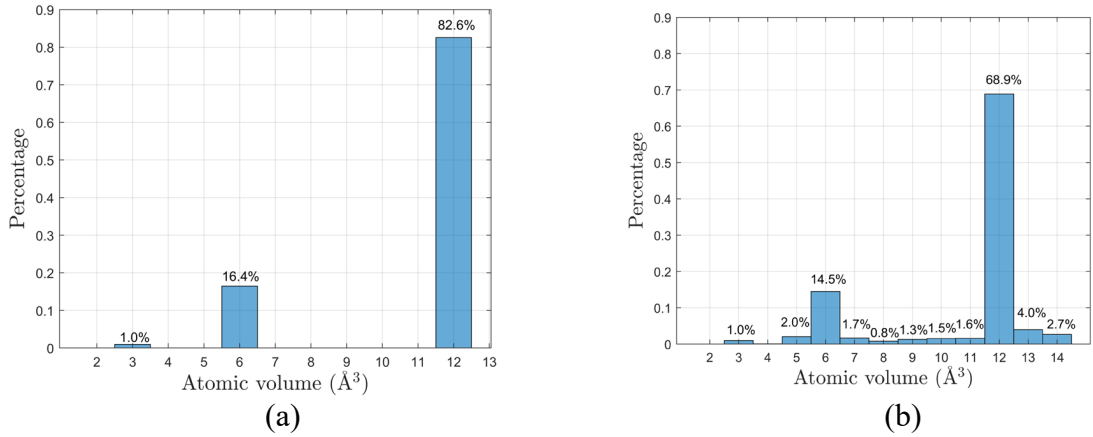

Fig. 21 The atomic volume distributions. (a) The distribution of the initial configuration and (b) the distribution of the 49th loading step configuration.

Figure 22 shows the stress distributions of the initial configuration and the 49th loading step configuration. Figure 22(a)(c) shows the stress distributions of the longitudinal sections of the two configurations. Figure 22(b)(d) shows the stress distributions of the cross-sections of the two configurations. In the initial configuration, the internal stresses were small and evenly distributed. The surface stresses were relatively large because the atomic coordination number of the surface atom was less than that of the internal atom. In the 49th loading step configuration, the internal stresses generally increased, but the distribution was no longer uniform. The stresses were concentrated near the defects because the original lattice structure was destroyed (defects were generated), and the strain energy density near the defects was higher than that in other regions. The stress distributions corresponded to the positions of different crystalline structures within the configurations. This result indicated that the stress distributions were consistent with the physical reality and that the atomic volumes calculated by our method were physically meaningful and correct.

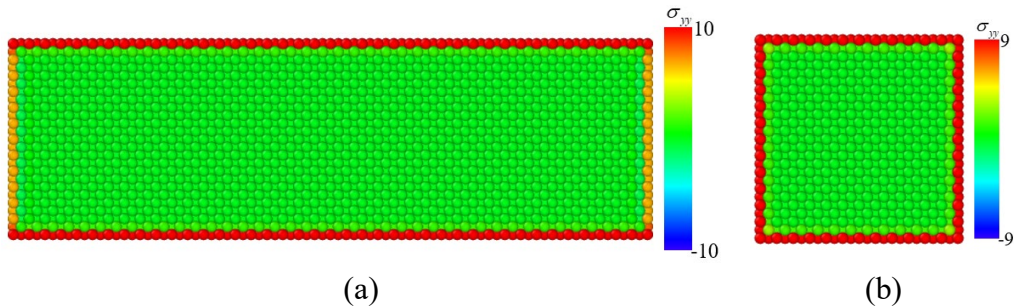

(a) (b)

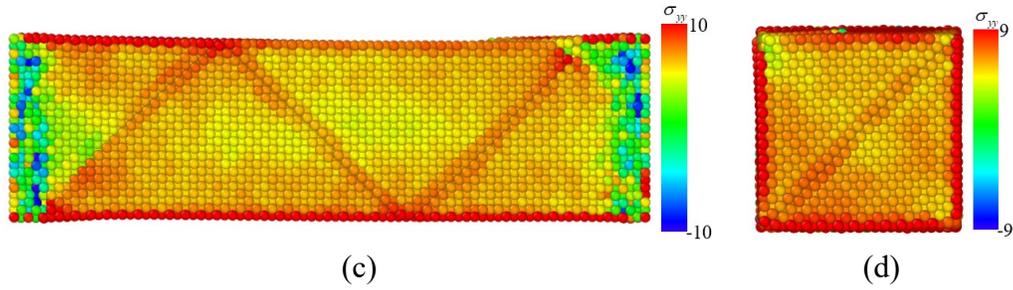

(c)  (d)

Fig. 22 Stress distributions (unit: GPa). (a) The distribution of the longitudinal section of the initial configuration (x<0), (b) the distribution of the cross-section of the initial configuration (y<0), (c) the distribution of the longitudinal section of the 49th loading step configuration (x<0), and (d) the distribution of the cross-section of the 49th loading step configuration (y<0).

*6.2 Cu nanowire bending*

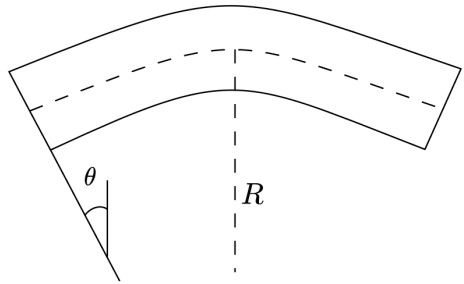

Fig. 23 A nanowire is bent at an angle $\theta$ and its radius of curvature is $R$.

A numerical example of bending of a Cu nanowire is provided in this subsection. The MD simulation of bending and unloading also used the EAM potential function. The initial configuration of the Cu nanowire is the same as that shown in Fig. 18. The bending process is shown in Fig. 23. An angle increment of one degree per loading step was exerted on both ends of the Cu nanowire. Specifically, all the atoms were moved to the corresponding positions according to the bending curvature radius $R$, which was determined by the bending angle $\theta$. Moreover, the center of the Cu nanowire was fixed, and the length of the neutral layer was kept constant during bending. After each loading step, the Cu nanowire was relaxed with its ends fixed at 300 K for 60 ps. The average position of each atom in the last 2 ps was recorded.

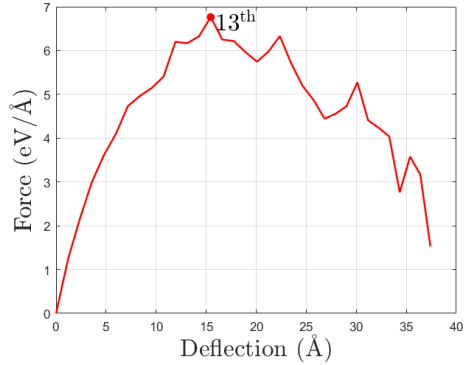

Fig. 24 The bending force–deflection curve of the single-crystal Cu nanowire.

Figure 24 is the bending force–deflection curve of the Cu nanowire. We chose the 13th loading step configuration to calculate the stress distribution. Figure 25 shows the longitudinal and cross-sections of the 13th loading step.

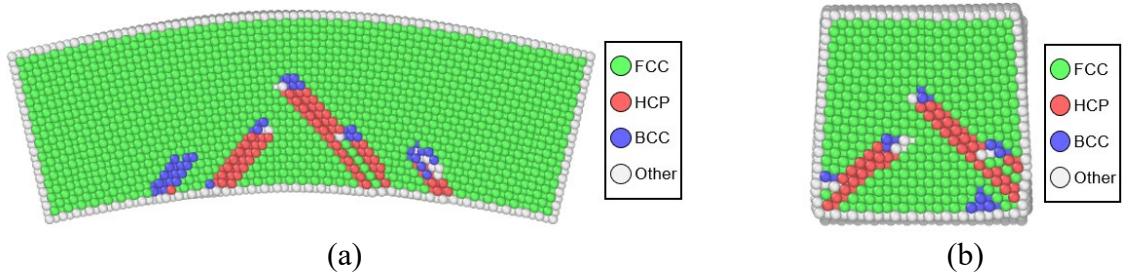

(a)          (b)

Fig. 25 The 13th loading step configuration. (a) The longitudinal section (x<0) and (b) the cross-section (y<0).

First, we calculated the atomic volume of the 13th loading step configuration, and Figure 26 shows the atomic volume distribution.

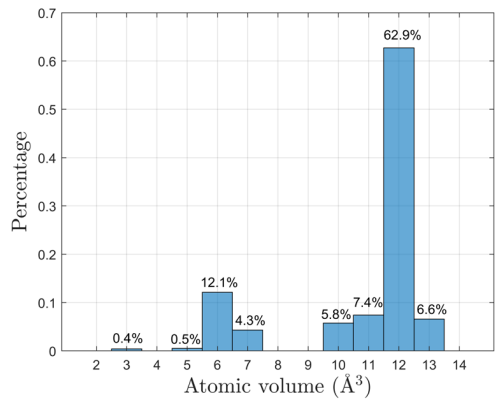

Fig. 26 The atomic volume distribution of the 13th loading step configuration.

Figure 27 is the stress distributions of the longitudinal and cross-sections of the 13th loading step. The stress distribution of the nanowire gradually increased along the positive y-axial direction, which corresponded to the curved shape of the nanowire. In addition, the stresses were relatively concentrated near the defects, which was caused by the destruction of the original crystalline lattice. The stress distribution of nanowire bending was consistent with the physical reality. This result indicated that the atomic volumes we calculated were physically meaningful and credible.

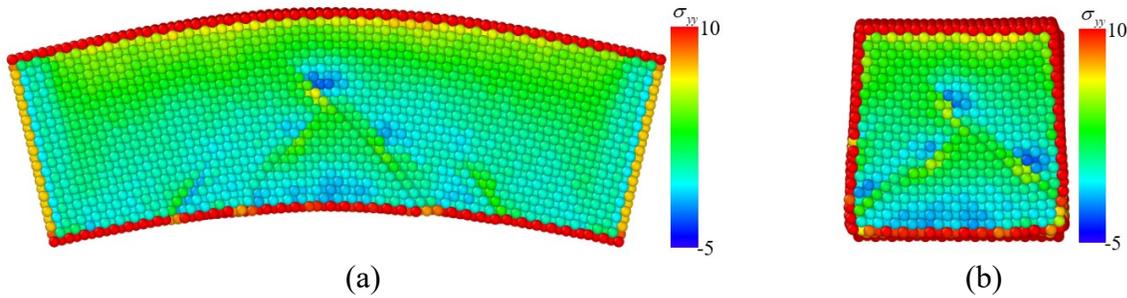

Fig. 27 The stress distributions of the 13th loading step configuration (unit: GPa). (a) The longitudinal section distribution (x<0) and (b) the cross-section distribution (y<0).

# 7 Conclusion

In multiscale research, researchers pay more attention to the calculation of macroscopic physical quantities from microscopic models. In many multiscale models, atomic volumes and surfaces are important innate quantities to calculate macroscopic physical quantities from atomic models. This paper proposes an atomic surface and volume calculation method for metal clusters based on the MS model. This method defines a variational Voronoi diagram. The variational Voronoi cell size is affected by the atom sphere size; the larger the atom is, the larger the cell. There are two atom types in the cluster: interior atoms and boundary atoms. For an interior atom, the method defines a variational Voronoi cell as its volume. For a boundary atom, the method defines the intersection of the overall cluster volume and its variational Voronoi cell as its volume. This atomic volume/surface method considers the effect of atom sizes. Therefore, the method is applicable not only to pure metals but also to alloys. The surface calculations of interior and boundary atoms are along with the volume calculations. Finally, we provide numerical examples of single-crystal Cu nanowire stretching and bending to verify our method. According to the numerical results, the stress distributions were consistent with the defect distributions within nanowires, which indicated that the atomic volumes/surfaces generated by our method are physically meaningful and accurate.

**Acknowledgments**

This work is supported by the National Natural Science Foundation of China [51739007] and the State Key Laboratory of Science and Engineering.

**Declarations of interest**

None.

**Data availability**

The raw/processed data required to reproduce these findings cannot be shared at this time as the data also forms part of an ongoing study.